\documentclass[showpacs,twocolumn,prb,superscriptaddress]{revtex4}
\usepackage{amsfonts}
\usepackage[latin9]{inputenc}
\usepackage{amsmath}
\usepackage{amssymb}

\usepackage{graphicx}

\DeclareMathOperator{\erf}{erf}

\begin{document}

\title{Thermodynamic properties of the electron gas in multilayer graphene in the presence of a perpendicular magnetic field}
\date{\today }
\author{B. Van Duppen}
\email{ben.vanduppen@uantwerpen.be}
\affiliation{Department of Physics, University of Antwerp, Groenenborgerlaan 171, B-2020
Antwerp, Belgium}
\author{F. M. Peeters}
\email{francois.peeters@uantwerpen.be}
\affiliation{Department of Physics, University of Antwerp, Groenenborgerlaan 171, B-2020
Antwerp, Belgium}

\begin{abstract}
The thermodynamic properties of the electron gas in multilayer graphene depend strongly on the number of layers and the type of stacking. Here we analyse how those properties change when we vary the number of layers for rhombohedral stacked multilayer graphene and compare our results with those from a conventional two dimensional electron gas. We show that the highly degenerate zero energy Landau level which is partly filled with electrons and partly with holes has a strong influence on the value of the different thermodynamic quantities.
\end{abstract}

\maketitle

\section{Introduction}

The relativistic character of the charge carriers in graphene has attracted a lot of interest. The unconventional quantum Hall effect\cite{Gusynin2005}, Klein tunnelling\cite{Katsnelson2006} and the Landau level spectrum\cite{Novoselov2005, Zhang2005} have shown that electrons in a layer of an hexagonal lattice of carbon atoms behave as two dimensional massless Dirac particles with a velocity 300 times smaller than the speed of light.

When several graphene layers are stacked on top of each other, the character of the charge carriers changes fundamentally with the number of layers and the type of stacking\cite{Partoens2007, Partoens2006}. The low energy behaviour of the electrons in multilayered structures can however be decomposed in a combination of multilayers with a lower number of rhombohedral stacked layers\cite{Partoens2006, Min2008}. Recent experimental progress proved that it is possible to fabricate multilayer samples with a specific number of layers and a specific type of stacking\cite{Shih2011}.  This resulted in an increased interest in the electronic properties of these multilayers.\cite{Nakamura2008, Koshino2010a, VanDuppen2013, VanGelderen2013} Experiments support the low energy theory for both bilayer\cite{Ohta2006a, Bao2010} and trilayer\cite{Zhang2011, Jhang2011, Taychatanapat2011} and results for other multilayers are expected soon. 

One of the most peculiar properties of a two dimensional electron gas (2DEG) is that upon the application of a perpendicular magnetic field, the energy spectrum is completely quantized and that several thermodynamic quantities like the Fermi level, magnetization and magnetic susceptibility have an oscillatory behaviour as a function of the magnetic field. \cite{Zawadzki1983, Zawadzki1984} This oscillatory behaviour has proven to be significantly different from the de Haas- van Alphen effect in three dimensional systems, indicating that it is a pure two dimensional effect. \cite{Eisenstein1985}

In this paper, we combine the two dimensionality of graphene with the relativistic character of the electrons to compare the thermodynamic quantities with those of a conventional 2DEG.  We investigate rhombohedral stacked multilayered systems using the two band approximation\cite{Min2008} and present analytical formulae for different thermodynamic quantities as a function of the number of layers for zero and non zero temperature. 

 We consider a two dimensional gas of non interacting electrons with only nearest neighbour interlayer and intralayer transitions. This allows us to present an analytical theory which can form the basis of a more in depth analysis that does include these corrections.\cite{Bao2010, Ohta2006a, Nilsson2006b, Elias2011, Kotov2012, Jia2013, Kretinin2013a}.

We find that our results are fundamentally different from those of a 2DEG due to the different Landau level spectrum and in particular because of the presence of a highly degenerate zero energy Landau level. The results however still show the vanishing magnetization at zero magnetic field, which is a signature of the two dimensionality of the system.

In the first two sections, Sec. \ref{Sec:Model} and Sec. \ref{Sec:LLquantization}, of the paper at hand we discuss respectively the electronic properties of graphene multilayers and the way the spectrum discretizes into Landau levels. Then we calculate the oscillations of the Fermi level, the magnetization and the magnetic susceptibility for zero temperature in Sec. \ref{Sec:ZeroT} and for finite temperature in Sec. \ref{Sec:FiniteT}.  In Sec \ref{Sec:Conclusion} we conclude the analysis with a summary and some remarks concerning many-body interactions and additional transitions.

\section{Electrons in graphene multilayers \label{Sec:Model}}

As discussed before\cite{Min2008}, the valence and conduction band in multilayer graphene touch each other in two inequivalent points in reciprocal space, the so called Dirac points. Therefore, the low energy behaviour of the charge carriers in graphene multilayers reside in the energy valleys near these two points. Because of the high energy barrier between both valleys, we consider them to be uncoupled so their presence can be solely incorporated in the degeneracy of the electron states. 

Near the Dirac point, the energy spectrum can be decomposed in non-interacting pseudospin doublets with chirality $N$. These pseudospin doublets have a similar low energy spectrum as that of a rhombohedrally stacked multilayer with $N$ layers. Its Hamiltonian can be approximated by
\begin{equation}
\hat{H}_{N}=\frac{v_{F}^{N}}{\gamma _{1}^{N-1}}\left[ 
\begin{array}{cc}
0 & \hat{\pi}^{N} \\ 
\left( \hat{\pi}^{\dag }\right) ^{N} & 0%
\end{array}%
\right] ,  \label{HamiltonianMultilayer}
\end{equation}%
where\cite{Wallace1947} $v_{F}\approx 10^{6}$ m/s is the Fermi velocity in monolayer graphene,  $\hat{\pi}=\hat{p}_{x}-i\hat{p}_{y}$ with $\vec{p}=\left( p_{x},p_{y}\right) $ the in-plane momentum and \cite{Partoens2006} $\gamma _{1}\approx 0.4$eV is the interlayer hopping parameter. Note that we have omitted the minus sign in front of $\gamma _{1}$ due to electron-hole symmetry. The corresponding dispersion relation is 
\begin{equation}
\varepsilon =\pm \frac{v_{F}^{N}}{\gamma _{1}^{N}}p^{N}.
\end{equation}%
The energy, $E$, is here expressed in units of the interlayer hopping parameter, i.e. $\varepsilon =E/\gamma _{1}$.

The two-band approximation neglects the skew hopping parameters\cite{Partoens2006} $\gamma _{3}\approx 0.29$eV and $\gamma _{4}\approx 0.12$eV that give rise to trigonal warping and to a violation of electron-hole symmetry\cite{Mucha-Kruczynski2010} which becomes only visible for large energy, i.e. $E>1eV$. Also the $\gamma _{2}\approx 0.02$eV and $\gamma _{5}\approx 0.02$ eV parameters are neglected because they correspond to next-to-nearest-neighbour interlayer transitions. The validity of the two-band approximation is therefore limited to energies $\left\vert \gamma _{2}\right\vert <E<\gamma _{1}$. However, we can take Eq. $\left( \ref{HamiltonianMultilayer}\right) $ also as a model Hamiltonian which allows us to obtain many results analytically. 

\begin{figure}[tb]
\centering
\includegraphics[width= 8.5cm]{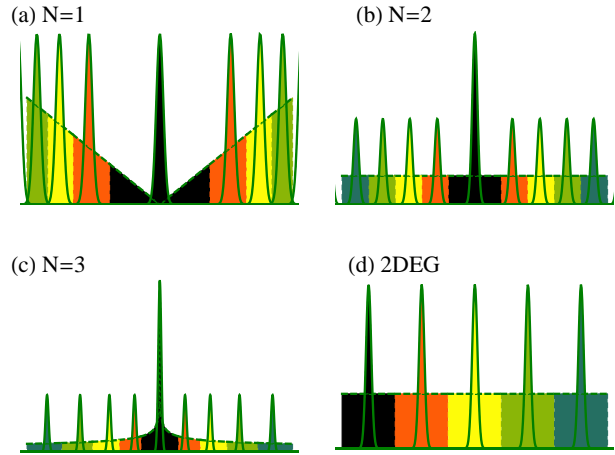}
\caption{(Colour online) Schematic representation of the DOS as a function of the energy of (a) monolayer, (b) bilayer and (c) trilayer graphene. (d) Results for a normal 2DEG. The LLs are shown as gaussian peaks. The different colours indicate with which LL the states from the zero field DOS are associated.} \label{FigDOSSchem}
\end{figure}

The density of states (DOS) of the two dimensional electron gas in multilayer graphene depends strongly on the power law of the dispersion relation. This is a big difference with respect to that of the normal 2DEG. The DOS per unit area $A$ is given by
\begin{equation}
\frac{D\left( \varepsilon \right) }{A}=\frac{2}{2\pi }\frac{2}{N}\frac{%
\gamma _{1}}{\left( \hbar v_{F}\right) ^{2}}\left\vert \varepsilon
\right\vert ^{\frac{2}{N}-1}.
\label{Eq:ZeroFieldDOS}
\end{equation}
This expression incorporates the extra valley degeneracy as the additional factor $2$. In Fig. \ref{FigDOSSchem} the DOS for zero and non zero magnetic field is shown for multilayers with $N$ up to $3$ and compared with those of the normal 2DEG.

\section{Landau level quantization}\label{Sec:LLquantization}

Similar to the 2DEG, the electronic states  of multilayer graphene discretize upon the application of a perpendicular magnetic field  leading to a quantized DOS into Landau levels (LLs)\cite{Guinea2006, Koshino2011, Pereira2007, Yuan2011a, VanElferen2013}. The magnetic field is incorporated in the Hamiltonian of Eq. $\left( \ref{HamiltonianMultilayer}\right) $ by the Peierls substitution $\vec{p}\rightarrow \vec{p}+e\vec{A}$. Using the Landau gauge $\vec{A}=B\left( 0,x,0\right) $ for convenience, this changes the operator $\hat{\pi}$ to $\hat{\pi}=\hat{p}_{x}-i\hat{p} _{y}-ieB\hat{x}$ which behaves now as a ladder operator similar to the case of a harmonic oscillator. Defining the lowering operator $\hat{a}=\frac{l_{B}}{ \sqrt{2}\hbar }\hat{\pi}$ and the raising operator $\hat{a}^{\dag }=\frac{l_{B}}{\sqrt{2}\hbar }\hat{\pi}^{\dag }$, with the magnetic length $l_{B}=\sqrt{\frac{\hbar }{eB}}$, the Hamiltonian from Eq. $\left( \ref{HamiltonianMultilayer}\right) $ can be written as

\begin{equation}
\hat{H}_{N}\left( B\right) =\gamma _{1}\alpha ^{N}B^{N/2}\left[ 
\begin{array}{cc}
0 & \hat{a}^{N} \\ 
\left( \hat{a}^{\dag }\right) ^{N} & 0%
\end{array}%
\right] ,
\end{equation}%
with $\alpha =\frac{\sqrt{2e\hbar }v_{F}}{\gamma _{1}}\approx 0.1/\sqrt{T }$, where $T$ stands for ``Tesla", the unit of magnetic field strength and the commutator $\left[ \hat{a},\hat{a}^{\dag }\right] =1$ ensuring a proper normalization of the ladder operators.

The eigenvalues and eigenstates of this Hamiltonian are found by solving the eigenvalue equation  $\hat{H}_{N}\Psi _{m}=E_{m}\Psi _{m}$ with the two-spinor 

\begin{equation}
\Psi _{m}=\left( 
\begin{array}{c}
\phi _{m} \\ 
\psi _{m}%
\end{array}%
\right) .
\end{equation}

The components of this two-spinor correspond to the atomic orbitals of the two free standing sublattices at the top and bottom layer. These two sublattices are the only two that do not lie directly below or above another sublattice in a rhombohedral multilayer. For monolayer graphene, they are the two inequivalent sublattices that are responsible for the pseudospin properties of the electrons. Using this two-spinor, one obtains the set of equations
\begin{equation}
\left\{ 
\begin{array}{c}
\varepsilon _{m}\phi _{m}=\alpha ^{N}B^{N/2}\hat{a}^{N}\psi _{m} \\ 
\varepsilon _{m}\psi _{m}=\alpha ^{N}B^{N/2}\left( \hat{a}^{\dag }\right)
^{N}\phi _{m}%
\end{array}%
\right. ,
\end{equation}%
with the dimensionless energy $\varepsilon _{m}=E_{m}/\gamma _{1}$. The
energy is found by solving the equation%
\begin{equation}
\frac{\varepsilon _{m}^{2}}{\alpha ^{2N}B^{N}}\psi _{m}=\left( \hat{a}^{\dag
}\right) ^{N}\hat{a}^{N}\psi _{m}.
\end{equation}%
Therefore, the second component of the spinor is an eigenstate of the number operator $\hat{m}=\hat{a}^{\dag }\hat{a}$ and the energy is given in terms of the eigenvalues $m$ of the number operator as\cite{Koshino2009a}
\begin{equation}
\varepsilon _{m,N}^{\lambda }\left( B\right) =\lambda \alpha ^{N}B^{N/2}%
\sqrt{\frac{m!}{\left( m-N\right) !}}\text{ for }m\geq N,
\label{EnergyNoPrime}
\end{equation}%
where $\lambda =1$ for electrons and $\lambda =-1$ for holes. The eigenstates corresponding to these eigenenergies are
\begin{equation}
\Psi _{m,ky}^{\lambda }=\frac{1}{\sqrt{2}}\left( 
\begin{array}{c}
\left\vert m-N,k_{y}\right\rangle  \\ 
\lambda \left\vert m,k_{y}\right\rangle 
\end{array}%
\right) \text{ for }m\geq N,
\end{equation}%
where $\left\vert m,k_{y}\right\rangle $ corresponds to the eigenfunctions of the number operator $\hat{m}$ and are given in position representation as
\begin{equation}
\left\langle \vec{r}|m,k_{y}\right\rangle =A_{m}e^{-\xi ^{2}/2}H_{m}\left(
\xi \right) e^{ik_{y}y},  \label{PosRepLadderFunctions}
\end{equation}%
with $A_{m}=1/\sqrt{\sqrt{\pi }2^{m}m!l_{B}}$ its normalization, $\vec{r}=\left(x,y\right)$, $\xi=l_{B}k_{y}+x/l_{B}$ and $H_{m}\left( \xi \right) $ the Hermite polynomial of order $m \in \mathbb{N}$. In addition to this series of Landau levels (LL), there is a zero energy Landau level (ZELL) that is $N$ times as degenerate as the rest of the LLs. This level has the eigenstates
\begin{equation}
\Psi _{m,ky}^{0}=\left( 
\begin{array}{c}
0 \\ 
\left\vert m,k_{y}\right\rangle 
\end{array}%
\right) \text{ for }0\leq m<N,
\end{equation}%
which corresponds to eigenfunctions of Eq. $\left( \ref{PosRepLadderFunctions}\right) $ located on only one of the two sublattices. Note that when we consider the other Dirac point, the other sublattice is occupied with these zero energy states\cite{Goerbig2011, Neto2009}. The ZELL is half filled with electrons and half with holes\cite{Novoselov2005, Zhang2005} and therefore it gives rise to the unconventional quantum Hall effect which has been observed in graphene multilayer structures\cite{Min2008, McCann2006, Novoselov2006, Guinea2006, Koshino2011}. Due to its high degeneracy it has also attracted a lot of attention recently in the framework of fractional quantum Hall studies and other many body effects\cite{Yu2013a, Chakraborty2013, Bao2010}.

\begin{figure}[tb]
\centering
\includegraphics[width= 8.5cm]{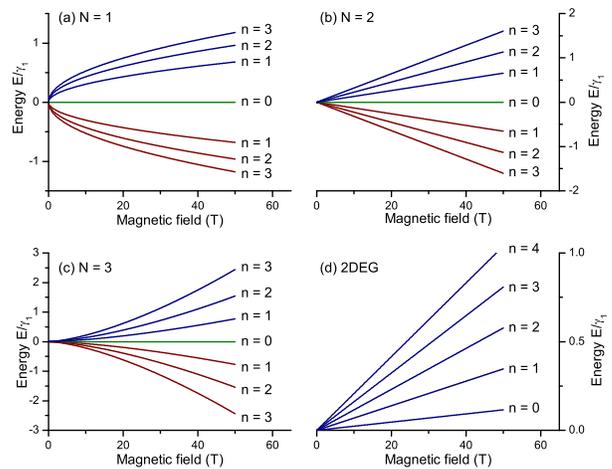}
\caption{(Colour online) Magnetic field dependence of the LLs in rhombohedral (a) monolayer, (b) bilayer and (c) trilayer graphene. (d) The LL spectrum for a 2DEG in GaAs with  $m^{*}=0.0665 m_{0}$ where $m_{0}$ is the free electron mass. The colour of the curve indicate the type of the charge carriers making up the LLs. Blue corresponds with electrons (positive energy), red with holes (negative energy) and green indicates the ZELL at zero energy. }
\label{FigSpectraLandau}
\end{figure}

Since we are interested in the thermodynamic properties of the electron gas in graphene multilayer structures, we renumber the LLs by $n=m-N+1$, so the energy spectrum changes into
\begin{equation}
\varepsilon _{n,N}^{\lambda }\left( B\right) =\lambda \alpha ^{N}B^{N/2}\sqrt{\frac{\left( n-1+N\right) !}{\left( n-1\right) !}}\text{ for }n\geq 0,
\label{EnergyRenumbered}
\end{equation}%
where the degeneracy of the $n=0$ LL is multiplied by a factor $N$ to account for the states corresponding to $m<N$. The magnetic field dependence of the LLs is shown in Fig. \ref{FigSpectraLandau} for various multilayer structures and compared with the spectrum of the 2DEG.

Note that in contrast to the 2DEG, the LLs of a graphene multilayer are not positioned at equidistant energy levels. However, the electron concentration at which the LLs are filled does scale linearly with the LL index $n$ for large $n$. They are therefore placed at equidistant levels of the electron concentration for large values of the LL index $n$.

\section{Zero temperature}\label{Sec:ZeroT}

\subsection{Fermi energy}

Due to the LL quantization, the Fermi level of the system will oscillate as the magnetic field pushes the LLs apart in a similar fashion as in the case of a normal 2DEG\cite{Zawadzki1983, Zawadzki1984}. The discretized DOS per unit surface area is given by
\begin{equation}
\rho _{N}\left( \varepsilon \right) =\frac{4}{2\pi l_{B}^{2}}\left[ N\delta
\left( \varepsilon \right) +\sum_{\lambda =\pm 1}\sum_{n=1}^{\infty }\delta
\left( \varepsilon -\varepsilon _{n,N}^{\lambda }\right) \right] ,
\label{DensityOfStatesAll}
\end{equation}%
where the LL energy $\varepsilon _{n,N}^{\lambda }$ is given by Eq. $\left( \ref{EnergyRenumbered}\right) $. The discretized DOS is schematically shown in Fig. \ref{FigDOSSchem} for mono- to trilayer structures and compared to the usual 2DEG. In this figure, the part of the DOS that will form a specific LL are coloured according to the colouring of the LL peak. Note that the degeneracy of all but the ZELL is $2/\pi l_{B}^{2}$, twice that of the 2DEG due to the additional valley degeneracy. The DOS given in Eq. $\left( \ref{DensityOfStatesAll}\right) $ covers both the electrons $ \left( \lambda =+1\right) $ and the holes\ $\left( \lambda =-1\right) $. In the following we will consider only electrons.

\begin{figure}[tb]
\centering
\includegraphics[width= 8.5cm]{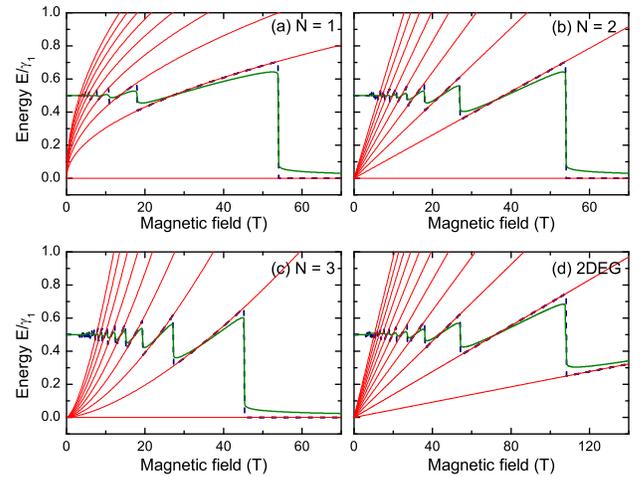}
\caption{(Colour online) Oscillatory behaviour of the Fermi energy as a function of the magnetic field. The green curve are the case of Gaussian LLs with width $\Gamma =0.05\protect\gamma _{1}$. The blue dashed line is for $\Gamma =0$ and the red thin curves are the LL spectrum for (a) mono-, (b) bi-, (c) trilayer graphene. (d) The Fermi level for a 2DEG in GaAs. The results are for an electron density such that the zero field Fermi level is at $E_{F,0}=0.5\protect\gamma _{1}$.}
\label{FigFermiFluctuation2}
\end{figure}

To calculate the Fermi level, $\varepsilon _{F}$, we assume the electron density $n_{0}$ to be independent of the strength of the applied magnetic field. The zero field Fermi level for a given concentration $n_{0}$ can be obtained using Eq. $\left(\ref{Eq:ZeroFieldDOS}\right)$
\begin{equation}
\varepsilon _{F,0}=\left( \frac{\hbar v_{F}}{\gamma _{1}}\right) ^{N}\left(
\pi n_{0}\right) ^{N/2}.
\end{equation}
For a normal 2DEG, the Fermi energy is proportional to the electron concentration. This is however not the case any more for multilayer graphene, where the number of layers determines the power of the relation. Using the discretized DOS from Eq. $\left(\ref{DensityOfStatesAll}\right)$ one obtains a relation between the electron concentration $n_{0}$ and the Fermi energy $\varepsilon_{F}$:
\begin{equation}
n_{0}=\frac{4B}{\phi _{0}} \sum_{n=0}^{\infty }g_{n}\theta \left( \varepsilon _{F}-\varepsilon
_{n,N}\left( B\right) \right)  ,
\label{FermilevelOsscilations}
\end{equation}%
where $\phi _{0}=h/e$ is the quantum of flux, $\theta\left(\ldots \right) $ is the Heaviside step function and the degeneracy of each LL is incorporated in the factor $g_{n}$ which is defined as
\begin{equation}
g_{n}=\left\{ 
\begin{array}{ccc}
N/2 & \text{if} & n=0 \\ 
1 & \text{if} & n>0%
\end{array}%
\right. ,
\end{equation}%
where the factor of $N/2$ is due to the aforementioned half occupancy of the ZELL with electrons. Solving Eq. $\left( \ref{FermilevelOsscilations}\right) $ for the Fermi level at a constant electron density results in a Fermi level that oscillates as a function of the magnetic field as shown in Fig. \ref{FigFermiFluctuation2} by the blue dashed curves.

Due to scattering or imperfections, the LLs are broadened. This can be incorporated by replacing the Dirac delta functions $\delta\left(\ldots\right)$ by a finite width Gaussian function given by
\begin{equation}
G_{n}\left( \varepsilon \right) =\sqrt{\frac{2}{\pi }}\frac{1}{\Gamma_{n} }\exp %
\left[ -2\left( \frac{\varepsilon -\varepsilon _{n,N}}{\tau_{n} }\right) ^{2}%
\right] ,
\end{equation}
where $\Gamma_{n} $ is the width of the $n^{th}$ LL and $\tau_{n} =\Gamma_{n} /\gamma _{1}$ its reduced value. Although the width may be different for each level, for convenience we will present numerical results for $\Gamma_{n}=\Gamma$, independent of the LL index. The electron concentration is
\begin{equation}
n_{0}=\frac{4B}{\phi _{0}} \sum_{n=0}^{\infty } \frac{g_{n}}{2} \erf\left( \sqrt{2}\frac{\varepsilon _{F}-\varepsilon _{n,N}}{\tau_{n} }\right)  ,
\end{equation}
where $\erf\left(\ldots\right)$ is the error function. The Fermi energy as a function of the magnetic field obtained using the above is shown in Fig. \ref{FigFermiFluctuation2} as solid green curves.

\begin{figure}[tb]
\centering
\includegraphics[width= 8.5cm]{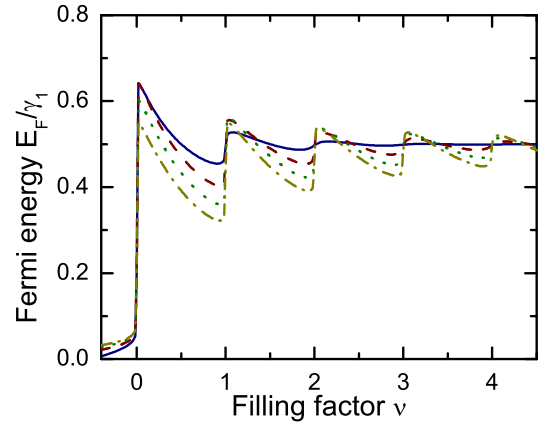}
\caption{(Colour online) The Fermi level as a function of the filling factor $\protect\nu $ for $N=1$ (solid blue), $N=2$ (dashed red), $N=3$ (dotted green) and $N=4$ (dash-dotted yellow) using the same parameters as Fig. \ref{FigFermiFluctuation2} with broadened LLs.}
\label{FigFermiFluctuationFillFac}
\end{figure}

The Fermi level converges to zero energy at increasing magnetic field because then all electrons are pushed in the ZELL. The transition to the $n^{th}$ LL occurs at the magnetic field $B_{n }$
given by
\begin{equation}
B_{n }=\frac{\phi _{0}n_{0}}{4\left( n +\frac{N}{2}\right) },
\label{BFieldFilling}
\end{equation}
where $n$ is a positive integer. This is a distinctive feature that differs from the behaviour of the 2DEG for which $B_{n}=\phi _{0}n_{0}/2n$ and $n>0$.  Note that since this quantity depends on the number of layers, $N$, the measurement of the magnetic field at which the magnetization changes rapidly can be used to determine the number of layers of the rhombohedral sample.

Using the expression of Eq. $\left( \ref{BFieldFilling}\right) $, we define the filling factor $\nu $ as a function of the magnetic field as
\begin{equation}
\nu \left( B\right) =\frac{\phi _{0}n_{0}}{4B}-\frac{N}{2}.
\end{equation}%
When the ZELL is completely filled and the other levels are empty, this filling factor is exactly zero. It can however also be smaller than zero, then the ZELL is only partly filled. Its lowest value, $-N/2$, is obtained for an infinite magnetic field. The filling factor can be decomposed into $\nu=p+\xi $ with
\begin{equation}
p=\left\lfloor \frac{\phi _{0}n_{0}}{4B}-\frac{N}{2}\right\rfloor \text{ and 
}\xi =\frac{\phi _{0}n_{0}}{4B}-\frac{N}{2}-\left\lfloor \frac{\phi _{0}n_{0}%
}{4B}-\frac{N}{2}\right\rfloor ,
\end{equation}%
where $\left\lfloor x\right\rfloor $ corresponds to the largest integer smaller than $x$, the highest fully occupied LL is $p$ and $\xi $ measures the partial occupation of the next LL and is a positive number smaller than one. Therefore, the Fermi level for non broadend LLs at $T=0$ is given by 
\begin{equation}
\varepsilon _{F} \left( B \right)=\varepsilon _{p\left( B \right)+1,N}.
\end{equation}%
This corresponds to an oscillating function that jumps between the different branches of the spectrum shown in Fig. \ref{FigFermiFluctuation2}. Since the interlevel transitions occur for different multilayers at different magnetic fields as given by Eq. $\left( \ref{BFieldFilling}\right) $, the investigation of the electronic properties as a function of the filling factor allows for a better comparison of different multilayers as shown in Fig. \ref{FigFermiFluctuationFillFac}.

The unconventional integer quantum Hall effect in multilayer graphene was shown earlier to give rise to plateaux in the Hall conductivities with value\cite{Min2008}
\begin{equation}
\sigma _{xy}=\pm \frac{4e^{2}}{h}\left( \frac{N}{2}+n \right) .
\end{equation}%
This agrees with the values of the magnetic field at which a new Landau level is started to be filled in Eq. $\left( \ref{BFieldFilling}\right) $.

\subsection{Magnetization and susceptibility}

\begin{figure}[tb]
\centering
\includegraphics[width= 8.5cm]{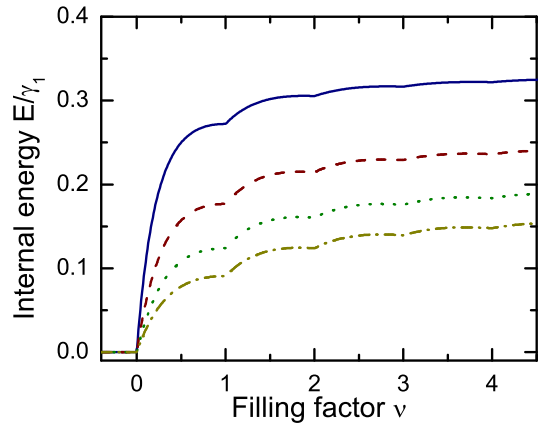}
\caption{(Colour online) Internal energy per electron for graphene multilayers with $N=1$ (solid blue), $N=2$ (dashed red), $N=3$ (dotted green) and $N=4$ (dash-dotted yellow). The electron concentration is such that the zero field Fermi level is $E_{F,0}=0.5\protect\gamma _{1}$. The width of the LL is $\Gamma=0$.}
\label{FigIntE}
\end{figure}

At $T=0$, the internal energy is generated by the first $p$ occupied LLs and a partial contribution from LL $p+1$. The internal energy per electron is therefore given by
\begin{equation}
u_{N}=\frac{4B}{\phi _{0}n_{0}}\left[ \sum_{n=1}^{p}\varepsilon _{n,N}+\xi
\varepsilon _{p+1,N}\right] ,
\end{equation}%
where $u_{N}=U/N_{0}\gamma _{1}$, the internal energy of $N$-multilayer graphene per electron, for a total number of $N_{0}$ electrons, and $\varepsilon _{n,N}$ is the single particle electron energy from Eq. $\left( \ref{EnergyRenumbered}\right) $. In Fig. \ref{FigIntE} we show the internal energy calculated as a function of the filling factor for various multilayers.

With increasing magnetic field, or decreasing filling factor, the occupation of the ZELL increases. Since the total number of electrons is kept constant, the internal energy per electron decreases at high magnetic fields because the electrons in the ZELL do not contribute to the internal energy. Therefore, when the field $B$ is larger than $B_{0}$, so $\nu <0$, $u_{N}$ remains constant at zero energy. This is another distinct feature from the normal 2DEG where at high magnetic field the lowest LL still contributes to the internal energy which increases with magnetic field. At zero field, the internal energy per electron $u_{N,0}$ will be 
\begin{equation}
u_{N,0}=\frac{\varepsilon _{F,0}}{\left( N/2+1\right) },
\end{equation}%
which also depends on the number of layers, $N$, of the system.

The magnetization is found by differentiating the free energy with respect to the magnetic field $M=-\partial F/\partial B$. Because at $T=0$ the free energy equals the internal energy, the magnetization per electron becomes
\begin{equation}
m_{N}=\frac{1}{B}\left[ \varepsilon _{p+1,N}-u_{N}\right] ,
\end{equation}%
where $m_{N}=M/N_{0}\gamma _{1}$ is the magnetization per electron of $N$-multilayer graphene and $u_{N}$ is the previously calculated internal energy per electron.

\begin{figure*}[tb]
\centering
\includegraphics[width= 17.5cm]{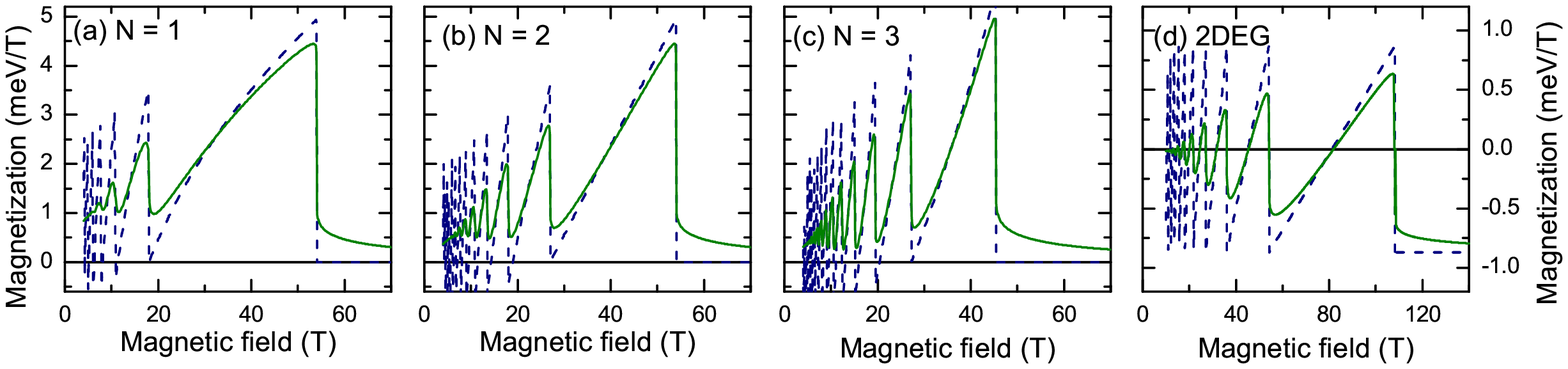}
\caption{(Colour online) Magnetization per electron as a function of the magnetic field for (a) monolayer, (b) bilayer, (c) trilayer graphene, and (d) 2DEG in GaAs. The dashed blue curves correspond to $\Gamma =0$ while for the green solid curve $\Gamma=0.05\gamma_{1}$.}
\label{FigMagnCol}
\end{figure*}

\begin{figure}[tb]
\centering
\includegraphics[width= 8.5cm]{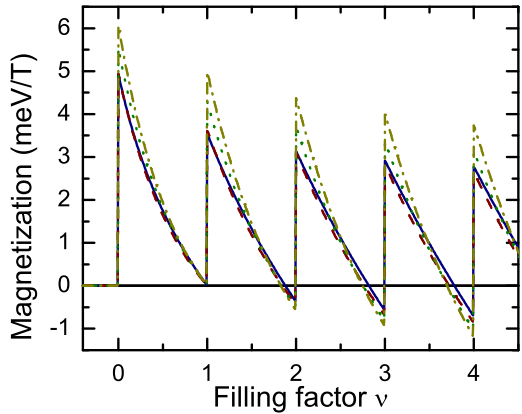}
\caption{(Colour online) Magnetization per electron for multilayers with $N=1$ (solid blue), $N=2$ (dashed red), $N=3$ (dotted green) and $N=4$ (dash-dotted yellow) as a function of the filling factor $\protect\nu $. The electron concentration is such that the zero field Fermi level is at $E_{F,0}=0.5\protect\gamma _{1}$. The width of the LL is $\Gamma=0$.}
\label{FigMagnetizationTZero}
\end{figure}

In Fig. \ref{FigMagnCol} we show the magnetization for different multilayers up to $N=3$ and compare it with the 2DEG. The sawtooth behaviour is similar to that of a 2DEG although it increases with magnetic field and becomes completely positive at large magnetic field before exponentially decreasing to zero for $B>B_{0}$. The tendency for positive magnetization is the consequence of the ZELL that does not contribute to the magnetization and is therefore important for large magnetic fields. For small fields, the magnetization oscillates around zero but keeps decreasing in magnitude. This is reminiscent of the non equidistant energy spacing of the LLs. Furthermore, only the bilayer case has a linear sawtooth magnetization because for a single LL the field dependence is $m_{N}\sim B^{N/2}$. When all the electrons occupy the zero energy LL, the magnetization is equal to zero since the internal energy does not change with the magnetic field anymore. The latter is distinctive from the 2DEG where the magnetization retains a finite constant value for large fields as shown in Fig. \ref{FigMagnCol}(d). In Fig \ref{FigMagnetizationTZero}, the magnetization is shown as a function of the filling factor for various multilayers. 

\begin{figure}[b!]
%\centering
\includegraphics[width= 8.5cm]{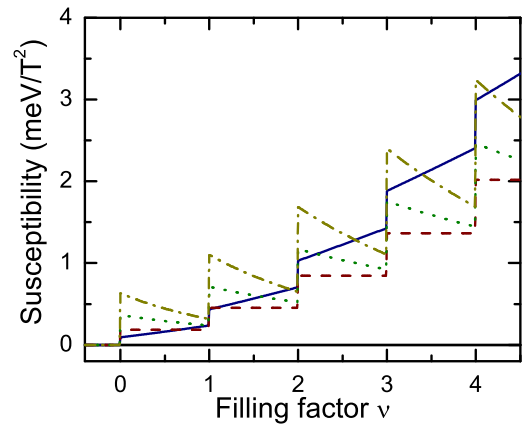}
\caption{(Colour online) Susceptibility of graphene multilayers with $N=1$ (solid blue), $N=2$ (dashed red), $N=3$ (dotted green) and $N=4$ (dash-dotted yellow) as a function of the filling factor $\protect\nu $. }
\label{FigSuscTZero}
\end{figure}

Differentiating once more, the susceptibility per electron $\chi_{N}=\partial m_{N}/\partial B$ is obtained
\begin{equation}
\chi _{N}=\frac{1}{B^{2}}\frac{N}{2}\left[ 2\varepsilon _{p+1,N}-u_{N}\right],
\end{equation}
where $u_{N}$ is the internal energy per electron. The susceptibility is shown in Fig. \ref{FigSuscTZero} as a function of the filling factor. Bilayer graphene has a constant step-like susceptibility similar to a 2DEG. 

\section{Finite temperature}\label{Sec:FiniteT}

\begin{figure}[tb]
\centering
\includegraphics[width= 8.5 cm]{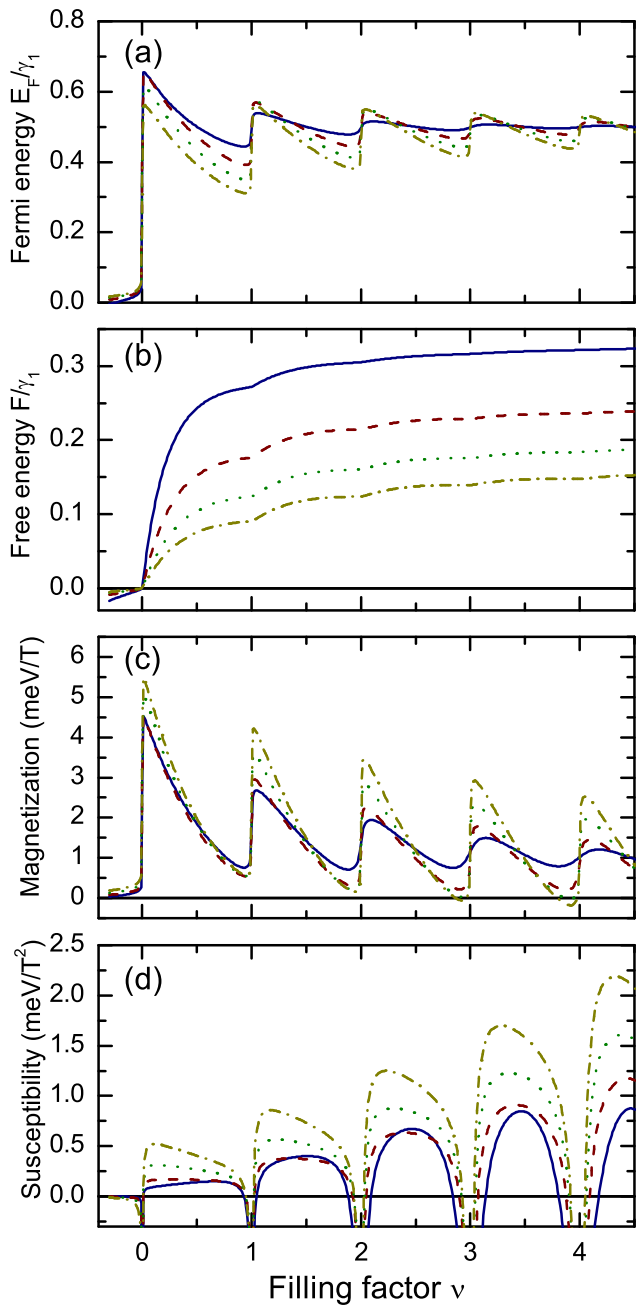}
\caption{(Colour online) Thermodynamic quantities at $T\approx 40K $ as a function of the filling factor for multilayer graphene with $N=1$ (solid blue), $N=2$ (dashed red), $N=3$ (dotted green) and $N=4$ (dash-dotted yellow). (a) Fermi level, (b) free energy, (c) magnetization and (d) magnetic susceptibility. }
\label{FigTNonZero}
\end{figure}

At finite temperature, the particle density for a spectrum of non-broadened LLs is given by
\begin{equation}
n_{0}=\frac{4B}{\phi _{0}}\sum_{n=1}^{\infty }g_{n}\left( 1+e^{\beta \left(
\varepsilon _{n,N}-\varepsilon _{F}\right) }\right) ^{-1},
\label{Eq:ElectronConcTNonZero}
\end{equation}%
where $\beta =\gamma _{1}/k_{B}T$ measures the inverse temperature and $g_{n} $ accounts for the degeneracy of each LL as before. The oscillations of the Fermi energy are damped as compared to the zero temperature result. This is shown in Fig. \ref{FigTNonZero}(a). 

At finite temperature, one needs to consider the free energy to find the thermodynamic quantities as the magnetization and the magnetic susceptibility. The free energy per unit area is\cite{Zawadzki1984}%
\begin{equation}
f_{N}=n_{0}\varepsilon _{F}-\frac{4B}{\phi _{0}\beta }\sum_{n=0}^{\infty
}g_{n}\ln \left( 1+e^{\beta \left( \varepsilon _{F}-\varepsilon
_{n,N}\right) }\right) ,
\end{equation}%
where $f_{N}=F/\gamma _{1}$ and the Fermi energy, $\varepsilon_{F}$, is found by solving Eq. $\left(\ref{Eq:ElectronConcTNonZero}\right)$ for a constant electron density $n_{0}$. Using the free energy, the magnetization is readily obtained by calculating its derivative with respect to the magnetic field
\begin{align}
m_{N,T}=\frac{4 }{\beta \phi _{0}}& \sum_{n=0}^{\infty }g_{n}
\left[ \ln \left( 1+e^{\beta \left( \varepsilon _{F}-\varepsilon _{n,N}\right)
}\right) \right.  
\notag \\ 
&\left.-\frac{N}{2}\beta \varepsilon _{n,N} \left( 1+e^{\beta \left(
\varepsilon _{n,N}-\varepsilon _{F}\right) }\right) ^{-1}\right]
 ,
\end{align}%
and the susceptibility becomes

\begin{widetext}
\begin{equation}
\chi _{N,T}=\frac{4}{\phi _{0}}\sum_{n=0}^{\infty }g_{n}\left( 1+e^{\beta
\left( \varepsilon _{n,N}-\varepsilon _{F}\right) }\right) ^{-1}\left[ 
\left( 1-\frac{N}{2}\beta \varepsilon _{n,N}\left( 1+e^{\beta \left(
\varepsilon _{F}-\varepsilon _{n,N}\right) }\right) ^{-1}\right)  
\left( \frac{\partial \varepsilon _{F}}{\partial B}-\frac{N}{2}\frac{%
\varepsilon _{n,N}}{B}\right)  
-\frac{N^{2}}{4}\frac{\varepsilon _{n,N}}{B}%
\right] ,
\end{equation}%
\end{widetext}
where the derivative of the Fermi energy is obtained by differentiating the
expression for the electron concentration%
\begin{widetext}
\begin{equation}
\frac{\partial \varepsilon _{F}}{\partial B}=-\frac{4
\sum_{n=0}^{\infty }g_{n}\left( 1+e^{\beta \left( \varepsilon
_{n,N}-\varepsilon _{F}\right) }\right) ^{-1}\left( 1-\frac{N}{2}\beta
\varepsilon _{n,N}\left( 1+e^{\beta \left( \varepsilon _{F}-\varepsilon
_{n,N}\right) }\right) ^{-1}\right) }{\beta B \sum_{n=0}^{\infty }g_{n}\cosh ^{-2}%
\left[ \frac{\varepsilon _{n,N}-\varepsilon _{F}}{2}\beta \right] }.
\end{equation}
\end{widetext}

In Figs. \ref{FigTNonZero}(c,d) we show the magnetization and susceptibility at $T\approx 40K$ for mono-, bi-, tri- and tetralayered structures. As compared to the zero temperature case, the oscillations are damped, but the larger the number of layers, the weaker the damping is.

\section{Conclusion and remarks\label{Sec:Conclusion}}

 In this paper, we have calculated the thermodynamic quantities of the non interacting electron gas in multilayer rhombohedral graphene structures in a perpendicular magnetic field at zero and non zero temperature. Due to the discretization of the DOS, the Fermi level, magnetization and susceptibility oscillate as a function of the magnetic field. In contrast to a 2DEG, multilayer graphene has a highly degenerate ZELL which causes the magnetization and the susceptibility to tend towards zero for fields above a critical magnetic field. The value of this critical magnetic field can in principle be used to determine the number of layers of a rhombohedral sample. With a finite temperature analysis, we have shown that with increasing temperature, the oscillations are damped, but that this effect is less pronounced in samples with a higher number of layers. 

The results obtained in this paper will be affected by electron - electron interactions\cite{Jia2013}, the inclusion of additional inter- and intralayer transitions\cite{Kretinin2013a}, the occurrence of stacking boundaries\cite{Gorbachev2013} or other corrections. Therefore, the paper at hand provides a basis and reference point for these studies.

\section*{Acknowledgments}

The authors would like to thank C. De Beule for enlightening discussions. This work was supported by the European Science Foundation (ESF) under the EUROCORES Program Euro-GRAPHENE within the project CONGRAN, the Flemish Science Foundation (FWO-Vl) by an aspirant research grant to B. Van Duppen and the Methusalem Programme of the Flemish Government.

\end{document}